\documentclass[runningheads]{llncs}
\usepackage{multirow}
\usepackage[T1]{fontenc}
\usepackage{graphicx}
\usepackage{hyperref}
\usepackage{color}

\usepackage{calc}
\usepackage{amsmath,amssymb}
\usepackage{ifthen}
\usepackage[normalem]{ulem} 

\newboolean{showcomments}
\setboolean{showcomments}{true} 
\ifthenelse{\boolean{showcomments}}
  {
		\newcommand{\nbb}[2]{
		\fcolorbox{black}{yellow}{\bfseries\scriptsize#1}
		{$\blacktriangleright$\textcolor{blue}{\textit{#2}}$\blacktriangleleft$}
		}
		
		\newcommand{\remarks}[1]{\color{red}[#1]\color{black}}

		\newcommand{\del}[1]{\textcolor{red}{\sout{#1}}} 
		
		\newcommand{\removed}[1]{\cbstart\removedfragile{#1}\cbend{}}
		\newcommand{\removedfragile}[1]{{\color{red}{\sout{#1}}}{}}

  }
  {
		\newcommand{\nbb}[2]{}
		\newcommand{\remarks}[1]{}

		\newcommand{\del}[1]{} 
		
		\newcommand{\removed}[1]{} 
  		\newcommand{\removedfragile}[1]{}

  }

\newcommand{\ext}[1]{}

\begin{document}
\title{Model-based Generation of Attack-Fault Trees}
\titlerunning{Model-based Generation of Attack-Fault Trees}
%
\author{Raffaela Groner\inst{1}\orcidID{0000-0001-8744-9203} \and
Thomas Witte\inst{1}\orcidID{0000-0001-5391-7419} \and
Alexander Raschke\inst{1}\orcidID{0000-0002-6088-8393} \and 
Sophie Hirn\inst{1} \and 
Irdin Pekaric\inst{2,3}\orcidID{0000-0002-0706-3202} \and
Markus Frick\inst{2} \and
Matthias Tichy\inst{1}\orcidID{0000-0002-9067-3748} \and
Michael Felderer\inst{2,4,5}\orcidID{0000-0003-3818-4442}}
\authorrunning{R. Groner et al.}
%
\institute{Institute of Software Engineering and Programming Languages, Ulm University, Germany,
\email{\{firstname.lastname\}@uni-ulm.de} \and
Department of Computer Science, University of Innsbruck, Austria \\
\email{\{firstname.lastname\}@uibk.ac.at}\and
Department of Information Systems and Computer Science, University of Liechtenstein, Liechtenstein \and
Institute for Software Technology, German Aerospace Center (DLR), Germany \and
Department of Mathematics and Computer Science, University of Cologne, Germany} 
\maketitle              
\begin{abstract}
Joint safety and security analysis of cyber-physical systems is a necessary step to correctly capture inter-dependencies between these properties. 

Attack-Fault Trees represent a combination of dynamic Fault Trees and Attack Trees and can be used to model and model-check a holistic view on both safety and security. Manually creating a complete AFT for the whole system is, however, a daunting task. It needs to span multiple abstraction layers, e.g., abstract application architecture and data flow as well as system and library dependencies that are affected by various vulnerabilities.

We present an AFT generation tool-chain that facilitates this task using partial Fault and Attack Trees that are either manually created or mined from vulnerability databases. We semi-automatically create two system models that provide the necessary information to automatically combine these partial Fault and Attack Trees into complete AFTs using graph transformation rules. 

\end{abstract}
\section{Introduction}
\label{sec:introduction}

As cyber-physical systems (CPS) become more and more ubiquitous, safety and security analysis of such systems must take new and emerging problems into consideration. The proliferation of connected and smart devices, their interaction, and constantly changing software (e.\,g., due to over-the-air updates) leads to new safety and security problems. In particular, the increasing inter-connectivity of CPSs  has a significant impact on their security. The amount of reported vulnerabilities increases year by year\footnote{\url{https://www.cve.org/About/Metrics}}. Each vulnerability can cause a failure of (parts of) the system, which in turn may affect its safety. In consequence, safety and security analysis in isolation, and without consideration of the specific environment the system is used in, usually cannot capture inter-dependencies between these concerns easily. The heterogeneity of CPSs in terms of hardware, but especially in terms of operating system versions with different installed package versions, requires that the deployed CPSs are analyzed and constantly monitored during operation. 

Attack-Fault Trees (AFT) as a combination of Attack Trees and dynamic Fault Trees enables the joint analysis of safety and security properties in a single modeling formalism \cite{Stoelinga2,Stoelinga1}. The AFT model can then be checked using existing model-checking techniques, e.\,g., critical path analysis or calculation of failure rates and probabilities. Generating large AFTs for realistic systems by hand, however, is error-prone and infeasible. 
It is much easier for safety experts to model Fault Trees on a system level and for security experts to create Attack Trees for used components~\cite{ComponentBasedHazard}. 

These partial models are often on very different levels of abstraction. On the one side, a Fault Tree for a safety hazard ends on the level of logical system components and data channels. On the other side, vulnerabilities are reported on the level of packages and libraries and not on the level of components. The creation of an Attack Tree for a component thus requires intimate knowledge of the implementation of this component. Our approach attempts to bridge this gap of abstraction levels by deriving system dependency models from a running system and combining Fault Trees and Attack Trees into an AFT on this basis. Since we allow the extension of the generated models by manual models at each modeling level, we refer to our approach as semi-automatic. 

\begin{figure}[htbp]
\includegraphics[width=\textwidth]{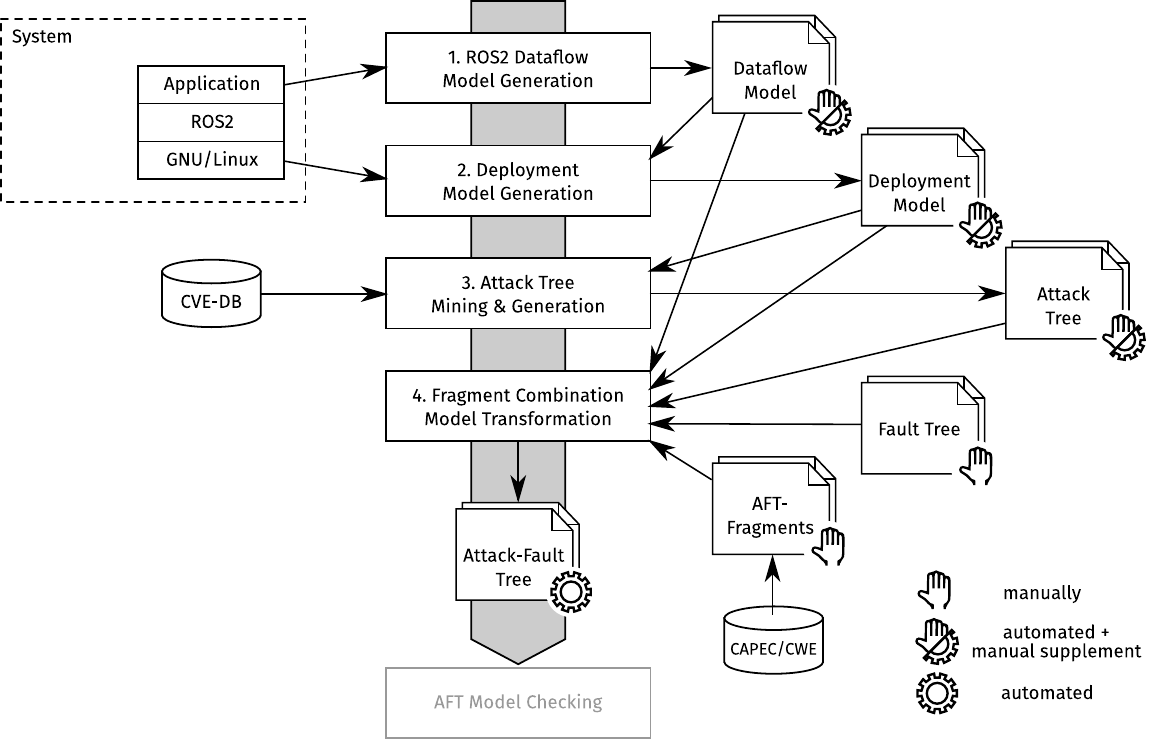}
\caption{Overview of the AFT generation toolchain and its models. Data sources for model generation are shown on the left and resulting models on the right.}
\label{fig:overview}
\end{figure}

In this paper, we present our toolchain (Fig.~\ref{fig:overview}) and preliminary evaluation for the semi-automatic generation of AFTs, previously outlined in \cite{witte2022towards}: As an input, we use Attack and Fault Trees created by security and safety experts respectively as well as additional Attack Trees, that are automatically mined from vulnerability databases for components, tools and libraries found on the analyzed system. Compared to \cite{witte2022towards}, which provided a vision of the approach, this paper provides a complete running pipeline. 

In order to combine these partial models into a single AFT, we use information from two system models. Separate models for the logical system architecture and dataflow \emph{(Dataflow Model)}, and the deployment and dependencies of these logical components \emph{(Deployment Model)} are automatically derived  from the running system and used to provide the necessary information to combine higher-level Fault Trees with system-level Attack Trees.

We then use graph transformation rules inferred from typical attack and fault propagation patterns to identify and generate missing AFT fragments to bridge the abstraction gap between the partial attack and Fault Trees. To reduce the number of candidate Attack Trees to attach to the partial Fault Trees, we annotate basic events with impact requirements for potential attacks. These requirements are then propagated and matched with the impact scores of basic attack steps derived from vulnerability databases.

We evaluate our implementation by analyzing a quadcopter system: the quadcopter pose is tracked using multiple cameras. Flight trajectories that avoid obstacles are generated on a PC and then sent to the quadcopter. Our toolchain connects to the system, automatically gathers dataflow and deployment information of the system, mines vulnerability databases for found library and system dependencies, and generates initial Attack Trees. Then it combines these models with manually created Fault Trees, using dataflow and deployment information to complete the Fault Trees to full AFTs that include potential system vulnerabilities. On the intentionally not updated operating system, our approach successfully found some possible attacks that might lead to the hazards modeled in the Fault Trees and created meaningful AFTs for this scenario. 

However, we identified some possible improvements especially with respect to the mapping of software packages to vulnerabilities in CVE-databases and related the precise decision, which of the generated Attack Trees can be applied to an AFT to minimize the amount of false positives. 

The outline of the paper is as follows: Sect.~\ref{sec:background} shortly introduces besides necessary background on safety and security models, security metrics, and the robot operation system, a running example used in the remaining paper. In Sect.~\ref{sec:contribution}, we present our approach to combined safety and security modeling and analysis followed by a discussion of its chances and limitations in Sect.~\ref{sec:discussion}. Sect.~\ref{sec:relatedwork} compares the presented approach to related ones before Sect.~\ref{sec:conclusion} concludes the paper. 

\section{Background}
\label{sec:background}
This section provides a short overview of safety and security models used in the following sections. We also provide a short overview of common vulnerability metrics and the robot operating system (ROS), the system platform our demonstrator focuses on. 

\subsection{Fault/Attack Trees and Their Combination}\label{sec:attackFaultTreeBackground}
\textbf{Fault Trees (FT)} are a popular model formalism from safety analysis that is used to model possible hazards and their causes~\cite{vesely1981fault}.
There are several variants of FTs~\cite{RUIJTERS201529}, but we will restrict ourselves to a model formalism including complex gates like SAND or PAND similar to dynamic FTs~\cite{DFTs} for modeling.

\textbf{Attack trees (AT)}~\cite{schneier1999modeling,mauw2006foundations} are models used to represent adversary actions regarding how a certain system or component can be targeted. Vertices present intermediate targets or attack steps (i.\ e., various types of exploits or vulnerabilities). Edges represent dependencies among the actions and (intermediate) targets. By introducing logical gates (AND, OR, PAND, SAND, etc.) more complex attack scenarios can be modeled. A path from a leave to the root of an AT is called an attack path~\cite{lallie2020review}. Moreover, weights can also be assigned to edges to include costs, probability, risks, or other metrics. 

Automatic generation of (simple) ATs is a complex task that can be achieved by employing various existing databases that include vulnerability, library, attack, and severity data.

\textbf{Attack-Fault Trees (AFT)}~\cite{AFTsOrig,Stoelinga1} integrate Attack Trees within Fault Trees. This is achieved by redefining the basic events of a FT which now describe not only accidental events (e.g., the sudden failure of a component) but also the failure of a component due to the malicious actions of an attacker. While ATs usually describe a general attack pattern, the Attack Trees attached to FTs must be directed at a specific target. Therefore, it is necessary to make the definition of ATs more specific for this purpose, as described in~\cite{AFTsOrig}.

In our approach, an attack event consists of a description of the attack, a reference to a model element from the deployment or the dataflow model (see Sect.~\ref{sec:dataflow}) and a requirement in the form of minimum CIA values (see Sect.~\ref{sec:vulnerabilityMetricsBackground}) that an attack on the linked component must possess in order to trigger the attack. Adapted from the external events from Vesely et al.~\cite{vesely1981fault}, we use a house shape to represent our attack events graphically.

\subsection{Vulnerability Metrics}
\label{sec:vulnerabilityMetricsBackground}
In the following, we discuss vulnerability-related terms that are utilized in the proposed modeling approach. CVE\footnote{Common Vulnerabilities and Exposures, \url{https://cve.mitre.org/}} data is used to uniquely distinguish different vulnerabilities. CWE\footnote{Common Weakness Enumeration, \url{https://cwe.mitre.org/}} entries represent specific higher-level groups to which CVEs are assigned in order to provide a hierarchy for vulnerability data. There exist two main hierarchies, which divide them into software and hardware weakness types. Most of the CVEs contain a CVSS\footnote{Common Vulnerability Scoring System, \url{https://www.first.org/cvss/}} vector. It provides various types of qualitative scores related to the severity of the vulnerability. Besides some more detailed information assessing the severity, it contains information about the impact of a CVE, the so-called CIA triad \cite{samonas2014cia}.

CPEs\footnote{Common Platform Enumerations, \url{https://nvd.nist.gov/products/cpe}} represent various systems, software, and platforms, which are represented using syntax for Uniform Resource Identifiers (URI) including a specific version of a software or library. This allows security engineers and researchers to exactly know which software is affected by a certain CVE. In cyber-security, attacks can also occur by exploiting multiple vulnerabilities, which form attack chains. Some of these chains are similar in a way that they address CVEs that belong to the same CWE or they have corresponding mechanics. In order to represent these similarities, CAPEC\footnote{Common Attack Pattern Enumeration and Classification, \url{https://capec.mitre.org/}} entries were created, which allow an easy understanding of common attacker actions. The aforementioned databases present a solid foundation for the proposed approach since they include vulnerabilities, weaknesses, platforms, and attack patterns.

\subsection{Robot Operating System (ROS)}

While our safety and security analysis toolchain is technology and system agnostic, we implemented specific dataflow and deployment generators for the \emph{Robot Operating System (ROS)}. ROS \cite{macenski2022robot} is a middleware for component-based robotic applications. It consists of various helper libraries, e.g., for message transport, standardized interfaces, and tools. Components, called \emph{Nodes}, communicate over named and typed channels (\emph{Topics}) and via RPC (\emph{Services}). Application components can use multiple implementation languages; 3rd party components provided by, e.g., hardware manufacturers might act as black boxes, showcasing heterogeneous systems in need of joint safety and security analysis of the system as a whole. 

\subsection{Running Example}
As a running example, we present the following scenario: an autonomous drone  might pose an injury hazard to a bystander in case of a collision. This can be caused through a mechanical malfunction causing the accident, or as a result of an attack on the control system of the drone.

The drone control system in our quadcopter lab consists of a camera array for optical tracking, that calculates the exact position and orientation of the drone at a high frequency. This pose data is then sent to a ROS application, which consists of several components that implement trajectory planning, obstacle avoidance, and position control, among others. The quadcopter is connected via WiFi and control commands are sent to it using the AR.Drone SDK, closing the position control loop.

We create a Fault Tree to model the injury hazard: if one or more components or channels in the position control loop fail, a drone operator standing near the drone might be injured. Refining this FT and figuring out which software vulnerabilities, exploits or weaknesses might be applicable to trigger these fault events is done using our proposed AFT generation toolchain. An excerpt of the generated AFT is shown in Fig.~\ref{fig:example_aft}. Two potential attacks that might lead to a failure of the position controller are identified and attached based on libraries used by this component.

\section{SafeSec Attack-Fault Tree Generation Toolchain}
\label{sec:contribution}
We developed the \emph{SafeSec Attack-Fault Tree Generation Toolchain (SAFT-GT)} (Fig.~\ref{fig:overview}) to semi-automatically create and analyze AFT models for self-adaptive systems. Our toolchain uses \emph{dataflow} (Sect.~\ref{sec:dataflow}), and \emph{deployment models}(Sect.~\ref{sec:deployment}) to capture the state of the system and uses this information to automatically combine generated \emph{Attack Trees} (Sect.~\ref{sec:atg}) and manually created \emph{Fault Trees} that use different abstraction levels. A set of \emph{combination rules} is used to find and connect these AFT fragments (Sect.~\ref{sec:AFTGeneration}). The complete toolchain including the models used for the running example can be downloaded here: \url{https://www.uni-ulm.de/in/sp/research/projects/safesec/}

\subsection{Dataflow Model}
\label{sec:dataflow}
Our dataflow model captures the logical components and dataflow of the system. We separate this logical view on the system from the actual implementation and deployment in the deployment model. The meta-model of the dataflow model is rather simple: it consists of \emph{components} -- entities that provide, transform or process data -- and \emph{channels} -- ways for components to communicate, send messages or observe the state of other components.

This high level of abstraction is necessary to easily map basic fault events to system components and channels. A safety engineer manually creating Fault Trees for the system needs to specify the origin of basic fault events by annotating respective components and channels. 

We designed the model to be simple to generate for systems using different middlewares or frameworks. Our model simplifies the ROS component meta-model similar to the abstract component meta-model in \cite{gherardi2013variability} in order to easily generate and integrate dataflow models from ROS systems as well as other systems. ROS nodes are mapped to components while ROS topics, services, and actions are mapped to channels. Additionally, the model is designed to be manually extensible: additional components and channels can be defined manually and interface with the rest of the model. For example, the camera system and infrastructure to optically track quadcopters might be manually added including an optical channel from the quadcopter to the camera components to model the optical tracking of the quadcopter position.

Our implementation includes a dataflow model generator for ROS2 systems. ROS nodes do not define their interface statically, but connect to topics and services dynamically. Therefore, our generator consists of a single ROS node, that can be triggered to collect architecture and dataflow data using ROS' introspection capabilities at runtime. Due to its design as a daemon that collects data at runtime, the generation of the dataflow model is fast and captures exactly the current state of the system. The generator can be triggered repeatedly in order to monitor the system for changes or architecture reconfigurations. Deriving the dataflow model from a static configuration/composition description instead would require complex static analysis and access to the source code of all nodes.

\subsection{Deployment Model}
\label{sec:deployment}
In order to bridge the gap between the high-level dataflow model and the components that are deployed on a certain system, we introduce a so-called deployment model. This model contains the information which component is running on which system. Our  toolchain automatically extends this initial information with the files and ultimately the libraries, a component depends on. Dependencies that cannot be derived automatically (e.g., because the platform a component runs on cannot be reached by our analysis tool) can be given manually. 

Unlike other existing tools, such as snyk\footnote{\url{https://snyk.io}}, we do not rely on component source code to obtain dependency information. Instead, our tool uses information about the open files of a component's running processes. 

\begin{figure}
    \centering
    \includegraphics[width=\textwidth-3cm]{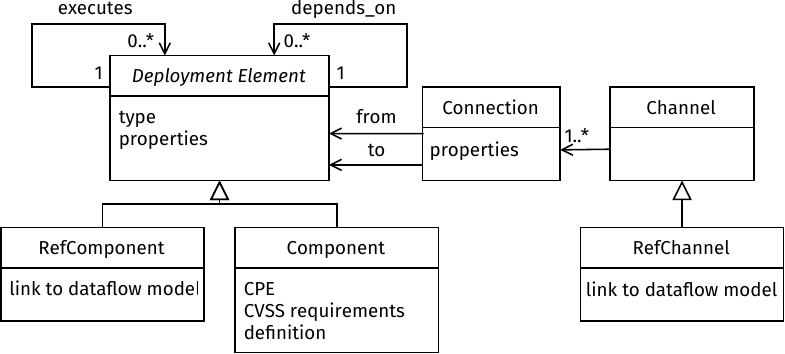}
    \caption{Simplified meta-model of deployment model.}
    \label{fig:deploymentmodel}
\end{figure}

Fig.~\ref{fig:deploymentmodel} shows a slightly simplified meta-model of the deployment model. Due to limited space, the \textit{Channel}s are shown but not described further below.  A \textit{Deployment Element} is either a newly defined \textit{Component} or a reference to a dataflow component (\textit{RefComponent}). Each component has a type (e.g., File, Library, Package, Platform, OS, HW, Sensor, Actor, ...) and arbitrary properties (key/value-pairs whose interpretation depends on the type). A (low level) component might have a CPE entry or CVSS requirements. If these requirements are met by an appropriate attack, the component can be considered as corrupted (see Sect.~\ref{sec:AFTGeneration}). 

Each deployment element can be executed on another element. In our running example most ROS nodes run on a PC called ``rosbox''. The properties of this ``rosbox'', including information how our analysis tool can reach this PC, are used to gather more detailed information about the components running on it. A deployment element can also depend on other elements. For instance, the component ``default\_FARFETCH\_bebop\_position\_control'' of our running example depends on library ``fast\_dds'' in version 2.1.1 (compare Fig.~\ref{fig:example_aft}). This dependency information is generated recursively by our analysis tool via the used files and libraries of a component returned by Unix tools like \texttt{lsof} and \texttt{ldd}. System specific package managers like \texttt{apt} and \texttt{dpkg} abstract this information into package names for which CVEs can be found. So far, the tool supports Ubuntu and Gentoo as platforms, but its architecture includes several abstraction layers to facilitate the integration of other platforms.

The next step is to find the corresponding CPE for each identified package. For this purpose, we use the tool CPEguesser\footnote{\url{https://github.com/cve-search/cpe-guesser}} in combination with some heuristic preprocessing like shortening names, removing additional version information, etc. Both lists, CPEs and all packages for which no CPE entry can be found, are then passed to the Attack Tree Generator to find possible CVEs for these pieces of software. 

\subsection{Attack Tree Generation}
\label{sec:atg}
The Attack Tree generator searches for vulnerabilities for a given set of software packages and generates (simple) ATs for each. Common CVE databases are utilized for this purpose. The selection of public information security databases was conducted based on the studies by Sauerwein et al. \cite{sauerwein2019analysis} and Pekaric et al. \cite{pekaric2021vulnerlizer}. As a result, NVD and MITRE databases were chosen as the most current and credible sources. For faster querying, all CVEs of these databases (including meta-information like CVSS) are cached in a 
local database. 

For CPEs, a specific query can be executed, while for general packages, a full-text search is performed. Once one or several CVEs are found for an entry, an Attack Tree containing all identified attack paths is generated using our self-defined DSL. 

In order to obtain a more extensive list of related CVEs, the CWE data, its hierarchy, and especially their relationship information like \textit{PeerOf}, \textit{CanFollow}, and \textit{CanPrecede} is considered in order to create attack chains in which multiple CVEs are linked using SAND, AND, and OR gates, telling the combination in which different CVEs must be exploited to conduct more complex attacks. Besides these automatically generated Attack Trees (ATs), more complex attack scenarios are only in a later step with the help of manually predefined AFT fragments. 

\subsection{Attack-Fault Tree Generation}\label{sec:AFTGeneration}

\begin{figure}[htbp]
    \centering
    \includegraphics[width=0.9\textwidth]{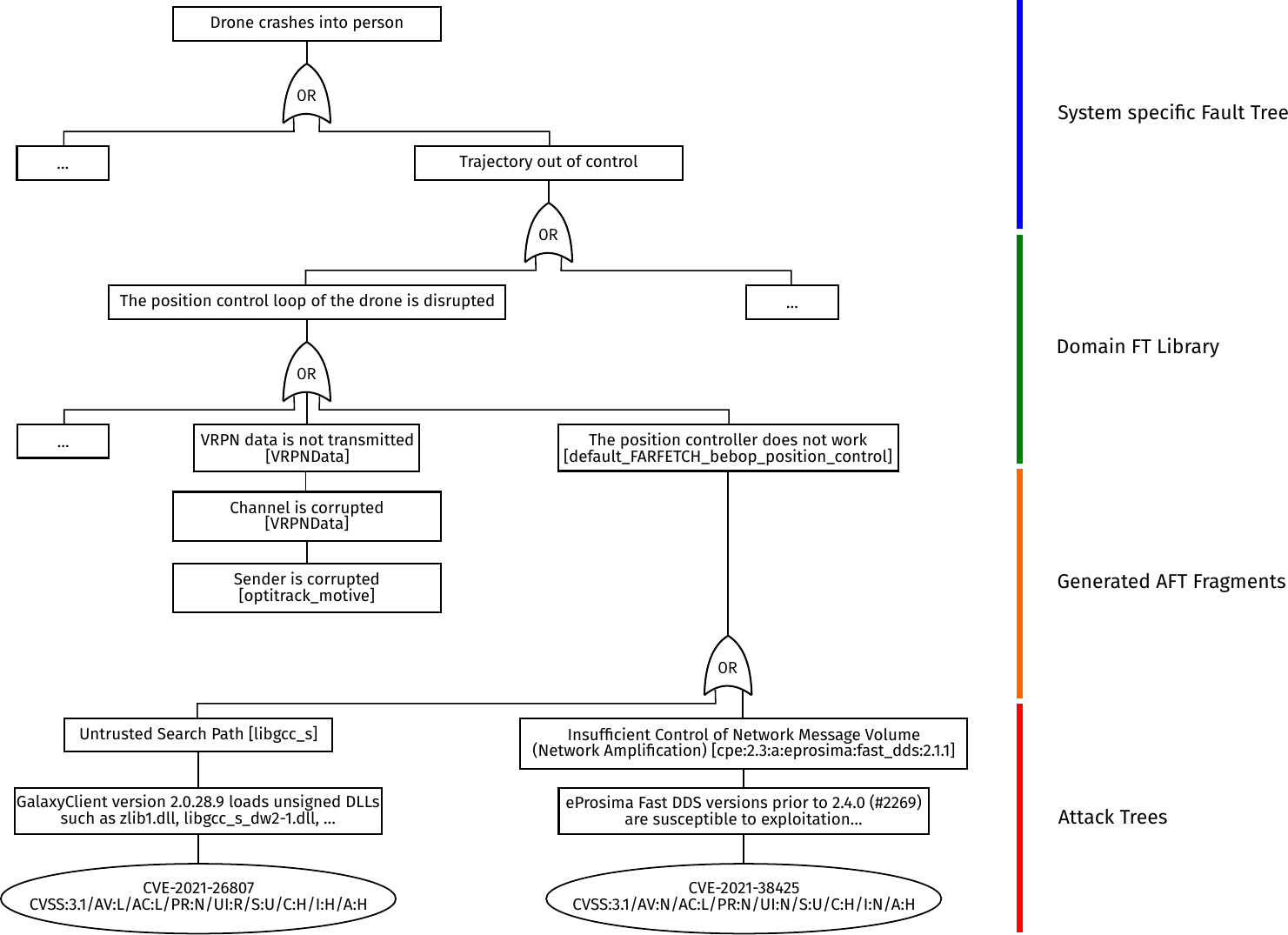}
    \caption{Generated AFT for the running example (truncated for readability).}
    \label{fig:example_aft}
\end{figure}
The AFT generation, as shown in Fig.~\ref{fig:overview}, mainly consists of combining fragments using model transformations based on our different input models. 
Overall, the AFT generation can be divided into three phases. First, 
the FT that forms the upper part of an AFT is copied into a new AFT model.
Second, the newly generated AFT model is extended using manually (pre-)defined AFT fragments, which represent generic, common attack patterns. Third,  appropriate ATs are attached at the leaves of the created AFT model.

To decide whether an AFT fragment or an AT can replace an attack event and thus be attached to the AFT we use the information about the system provided by the dataflow and the deployment model.
Since copying the FT into the AFT is trivial, we will present only the last two phases of AFT generation.

\noindent\textbf{AFT Fragments-phase:}  
The goal of the second phase is to create a bridge between the more abstract FTs, which are based on events caused by corrupted logical data flow or components, and the very technical ATs, which model, e.g., attacks on individual hardware or protocols.
This bridge is formed by our AFT fragments. 
These fragments represent different attack patterns that help to break down the more abstract attack events of the FT to the same level as the ATs.
Such a fragment describes for example the individual steps, which are necessary to perform an adversary in the middle attack (AiTM) or another one describes the relationship that when a sender is corrupted its associated channel on which it transmits is also corrupted.

We have defined two types of preconditions for each AFT fragment, which must be fulfilled, to replace an attack event in the AFT.
The first type of precondition defines the context of an attack event, namely the dataflow and/or components of the underlying system necessary to perform an attack.
The context of an attack event is defined by the referenced deployment model element or dataflow model element.
For example, an attack event must reference a channel from the dataflow model and the deployment model must define a communication over TCP/IP or UDP for the channel to be prone to an AiTM attack and thus allow to attach the respective fragment. 
In order to attach the AFT fragment describing the relationship between a corrupted sender and its channel, an attack event needs to reference a channel and there needs to be a component in the dataflow model which writes to this channel.

The second type of precondition is the expected impact of an AFT fragment to the confidentiality, the integrity and the availability (CIA triad) of the component or channel referenced by an attack event.
Each of the three aspects can take one of four values, namely * (any), L (low), N (neutral) and H (high).
A CIA value required by an attack event is satisfied if the attack described by an AFT fragment provides the same or a higher impact for a given aspect.  
For this we use the following order of the values: * $<$ L $<$ N $<$ H.
For example, the attack event ``VRPN data is not transmitted'' from our original FT defines that it can be replaced by an AFT fragment whose attack has a low impact on the confidentiality and a more than neutral impact on the integrity and availability.
Thus, even if the context is correct, an AiTM attack which aims to obtain data cannot replace this attack event since the confidentiality value of this AiTM attack is high and its integrity and availability values are low.
Our AFT fragment describing the relationship between a corrupted sender and its channel, defines that such an attack towards a sender has a high impact on the integrity and a neutral impact on the confidentiality and availability.
Thus, this AFT fragment satisfies both types of preconditions specified by the attack event ``VRPN data is not transmitted'' from our initial FT and is attached to the AFT, as shown in Fig.~\ref{fig:example_aft}.  

AFT fragments can also introduce new attack events, which then can be further replaced with other AFT fragments or ATs.
For example, ``Sender is corrupted'' in Fig.~\ref{fig:example_aft} is an attack event introduced by our AFT fragment describing the relationship between a corrupted sender and its channel.

In our proof of concept implementation, we manually defined five AFT fragments and their preconditions, which we have developed based on our expertise in modeling AFTs and ROS. 
We also analyzed the available list of CAPEC entries to identify entries that are suitable as a basis for AFT fragments and, e.g., used the CAPEC-94~\cite{CAPEC94} as basis for our AiTM AFT fragment.

\noindent\textbf{Append AT-phase:} In the last phase of AFT generation, we attempt to replace attack events that were not replaced in the previous phase or are newly added by the AFT fragments with generated ATs.
In order to decide whether an AT is suitable to replace an attack event or not we use the same two types of preconditions, namely the context and the CIA values defined in the respective attack event.

Since we generate the ATs, their preconditions regarding their context are much simpler than for the AFT fragments we define manually.
For an AT, the precondition for its context is already fulfilled if the context referenced by the attack event is mentioned in the description, the CPE or the name of the AT.
If an attack event references a deployment model element, it is checked whether this referenced element or one of its subcomponents is affected by the attack described in the AT.
If the attack event references a dataflow element, the deployment model is first searched for a more precise system description for the referenced dataflow model element.
If a corresponding element is found in the deployment model, it is checked as before whether this element or one of its subcomponents is affected by the attack described in the AT.

To decide whether the values of the CIA triad are fulfilled by the attack we use the corresponding values from the CVSS vector of the generated AT and the same rules as already used to attach our AFT fragments.
In our example, the attack event ``The position controller does not work'' defines that an attack must have at least one low impact on the confidentiality and at least a neutral impact on the integrity and availability.
This requirement is fulfilled by two generated ATs since according to its CVSS vector the AT ``Insufficient Control of Network Message Volume'' contains the value high for confidentiality and availability and a neutral one for the integrity. 
The CVSS vector for the AT ``Untrusted Search Path'' even possesses a value high for all three CIA values.
Since we rely on CIA values, our approach is limited to ATs that possess a CVSS vector.

\section{Discussion}
\label{sec:discussion}
We have modeled two different FTs to demonstrate the feasibility of our approach. 
The first one describes the possible injury of a person by a drone, we have also used this FT in our running example.
The second FT describes how a privacy violation by a drone can occur due to errors or malicious behavior.
We performed our AFT generation for both FTs in the context of our quadcopter lab. In total, the three phases of the AFT generation took approx. 15 s for the first and 10 s for the second FT on a Intel Core i7-3770 CPU with 16GB RAM.

Based on these results and a manual review of the generated AFTs, we can conclude that the whole pipeline, starting with the automated dataflow extraction of a (intentionally not updated) running ROS system generates correct and useful AFTs in a reasonable time. The most time consuming part is the search for vulnerabilities, especially in the case of full-text search. To improve the performance, the intermediate results could be cached and only recalculated, if new CVEs are reported and/or new package/file (versions) are detected. 

Unfortunately, some of the generated and attached attacks are false positives: The AT ``Untrusted Search Path'' fulfills all our preconditions, but the described attack does not apply to our system. 
The reason for this is that identifying software and software versions and mapping them to vulnerability data is hard. CPEs mitigate this problem by uniquely identifying affected software and software versions. However, to our knowledge, no mapping between arbitrary OS packages and CPEs exists and not all CVE entries contain affected CPE information. Our fallback to a full-text search may result in wrong or incomplete mappings due to different package names (e.g., software split into multiple packages, OS specific package naming), different versioning (e.g., due to additional applied patches), or renamed software projects. We only applied some simple heuristics, and thus the results can definitely be improved by using more sophisticated mining techniques such as NLP and/or by merging more detailed package databases. We think, this is an interesting research field that many other CVE-related research approaches could benefit from.

Moreover, our precondition consisting of the consideration of the three CIA values may be too vague to decide reliably whether an AFT fragment or an AT should be attached or not.
Here, a solution could be to include the other metrics provided by a CVSS vector in the decision.
In addition, we have defined only a few AFT fragments and we did not yet conduct a structured review of all CAPEC mechanisms and other related taxonomies of common attack patterns to derive and evaluate a complete set of general AFT fragments. We envision future iterations of our toolchain to provide a core set of general AFT fragments that can then be tailored toward specific domains and application areas by adding additional, more specific, AFT fragments.

\section{Related Work}
\label{sec:relatedwork}
The research field of combined security and safety analysis in software and systems engineering is huge~\cite{SLR}. For this reason, we will concentrate on the aspect of Attack/Fault Tree generation in our discussion of related work and only briefly address other aspects. 

The idea of generating fault- or attack graphs is not new. Swiler et al.~\cite{LauraSwiler2001} present a tool that generates attack graphs based on an assessment of security attributes and vulnerabilities in computer networks. Similar to our presented approach the authors use vulnerability scanning tools and attack templates. They mention the integration of an attacker profile might be interesting but also do not take this into account. Besides their focus on network attacks, the main difference to our approach is our attempt to combine the generated Attack Trees with Fault Trees. The configuration files introduced by Swiler et al. are similar to our dataflow and deployment models, but we generate these partially automatically. 

Kotenko et al.~\cite{Kotenko2013} utilize in their approach for Attack Tree generation similar techniques as proposed by us: CPEs are used to identify CVEs for components and CAPECs are employed for the generation of more complex attack scenarios. However, the main difference is their focus on network scenarios. Therefore, they use network security detection tools to identify possible vulnerabilities. Instead, we build upon software packages of a running (ROS) system. Similarly, Ou et al.~\cite{Ou2006} generate attack graphs for network topologies using logic programming. One disadvantage of their approach is that all information of the system must be given manually in advance by ``facts'' in the logical programming language. 

Our combination of FTs and ATs into AFTs follows the general approach stated by Steiner et al.~\cite{SteinerLiggesmeyer2016} which is in accordance with the work of Fovino et al.~\cite{AFTsOrig}. They describe the introduction of so-called ``security events'' in FTs in contrast to the existing ``safety events''. In contrast, Stoelinga et al.~\cite{Stoelinga1} introduce new model elements in order to combine FTs with ATs.

\section{Conclusion and Future Work}
\label{sec:conclusion}
In this paper, we presented our automated tool pipeline for generating AFTs based on generated and manually supplemented models. To bridge the gap between high level Fault Trees and low-level Attack Trees,  we introduce different intermediate models that describe the data flow between components and system dependencies. Using these models, we extend manually created FTs with generic and specific AFT fragments, and then attach generated ATs to the created AFT.
The advantage of this combined approach is the possibility to add manually created (partial) models at any stage. This allows this approach to be used even if the level of automation in a particular environment is not yet very high. 

At the moment, however, we see potential for improvement with respect to the mapping of used software packages and their vulnerabilities and the decision of whether an AT can be attached to the AFT or not. 
Here, we see several possibilities for improvement we plan to investigate in the future, from which also other research approaches can benefit. 
Also the extension of our approach towards other operating systems and software platforms besides ROS2 is an interesting future research direction. 

\subsubsection{Acknowledgements} This work was partially supported by the Austrian Science Fund (FWF): I 4701-N and the German Research Foundation (DFG): 435878599. 

%
%
%
\bibliographystyle{splncs04}
\bibliography{references}
\end{document}